\documentclass{article}
\usepackage{spconf,amsmath,epsfig}
\usepackage{amsfonts}
\usepackage{fancyhdr}
\usepackage{url}
\let\oldthebibliography\thebibliography
\renewcommand{\thebibliography}[1]{%
  \oldthebibliography{#1}%
  \setlength{\itemsep}{0pt}%
  \setlength{\parsep}{0pt}%
}

\fancypagestyle{ieeearxiv}{
  \fancyhf{}

  \fancyfoot[C]{%
    \parbox{0.96\textwidth}{%
    \centering
      \fontsize{6}{7}\selectfont

      \copyright\ 2019 IEEE.
      Personal use of this material is permitted.
      Permission from IEEE must be obtained for all other uses,
      in any current or future media, including
      reprinting/republishing this material for advertising or
      promotional purposes, creating new collective works, for
      resale or redistribution to servers or lists, or reuse of
      any copyrighted component of this work in other works.

      \par\vspace{2pt}

      This is the accepted version of the following paper:
      X. Xu, Q. Cai, J. Lin, S. Pan, and L. Ren,
      ``Enforcing Access Control in Distributed Version Control Systems,''
      in \textit{2019 IEEE International Conference on Multimedia and Expo
      (ICME)}, Shanghai, China, 2019, pp.~772--777,
      DOI: 10.1109/ICME.2019.00138.
    }%
  }
}

\begin{document}\sloppy

\def\x{{\mathbf x}}
\def\L{{\cal L}}

\title{Enforcing Access Control in Distributed Version Control Systems}
%
\name{Xin Xu\textsuperscript{1,2,3},
      Quanwei Cai\textsuperscript{1,2,*},
      Jingqiang Lin\textsuperscript{1,2,3},
      Shiran Pan\textsuperscript{1,2,3},
      Liangqin Ren\textsuperscript{1,2,3}\thanks{*Corresponding Author. This work was partially supported by Cyber Security Program of National Key R\&D Plan of China (Award No. 2017YFB0802100), and National Natural Science Foundation of China (Award No. 61772518).}}
\address{\textsuperscript{1}State Key Laboratory of Information Security, Institute of Information Engineering, CAS \\
\textsuperscript{2}Data Assurance and Communication Security Research Center, Chinese Academy of Sciences \\
\textsuperscript{3}School of Cyber Security, University of Chinese Academy of Sciences}

\maketitle
\thispagestyle{ieeearxiv}

\begin{abstract}
Version control systems (VCS), including central VCS (CVCS) and distributed VCS (DVCS),  are widely adopted to manage the changes to various types of data. 
Unlike the CVCS where all the entities obtain the data from the server and the access control is enforced with the cooperation of the server, each entity in the DVCS stores  the entire repository, obtains the repository shared by any entity and is free to share its own repository. 
Therefore, existing access control schemes for CVCS are not suitable for DVCS.
In this paper, we present a distributed access control scheme (Disac) for DVCS. Disac makes each entity have full control over its data, while the access control is enforced at each entity independently. We adopt  Attribute-based Encryption (ABE) and  Attribute-based Signature (ABS) to achieve the read and write permission control. The analysis of the Git client demonstrates that Disac can be integrated easily.
\end{abstract}


\section{Introduction}
\label{sec:intro}

The stand-alone or built-in version control systems (VCS) allow multiple entities to work on the same project concurrently, and are widely adopted to manage the changes to different types of data.
For example, Git \cite{git.com} provides services for source code, while Dropbox \cite{dropbox} and SharePoint Online \cite{sharepointonline} adopt the built-in VCS for the management of documents including images or videos, etc.

Existing VCS can be classified into central VCS (CVCS) and distributed VCS (DVCS): i) CVCS (e.g., CVS~\cite{cvs}, SVN~\cite{svn}) are based on the client-server architecture, in which a central server provides the reference version, and each client communicates with the server to synchronize its modifications. ii) DVCS (e.g., Git, veracity~\cite{veracity}) adopts the peer-to-peer approach, where each peer has a working directory and maintains its own repository comprising a full change history of the project, and synchronizes the repository by exchanging modifications from peer to peer.

Compared to CVCS, DVCS provides a more efficient and flexible cooperation, and avoids the single point of failure. In DVCS, each entity works on its local repository, and completes the common operations (e.g., commits of the modification, viewing history, and reverting changes) without any communication with others.
The entities are free to create, maintain or delete any local branch, which allows them to try a new idea or the one different from others privately.
Moreover, each entity chooses the modifications to share with others, and stores the copies (including the change-history) pulled from others, which provides a backup of the data and avoids relying on a certain entity for storing the data.

The repository may contain the sensitive information that cannot be read or modified by unauthorized entities (including the storage servers).
Various schemes are proposed to enforce access control  for cloud storage, which are suitable for the central VCS, but cannot be applied in the DVCS.
These schemes either fail to  manage the write permissions~\cite{Ateniese2006Improved,Vimercati2007Over,gitcrypt,git-remote-gcrypt} or rely on the cooperation of the single central server~\cite{Ateniese2006Improved,Sabrina2013Enforcing,Ruj2012Privacy,Zhao2011Realizing}.
For example,  the servers need to perform the re-encryption for each delegated user in the schemes~\cite{Ateniese2006Improved} based on proxy re-encryption, verify whether the modifier has the knowledge of the corresponding write `tag' in the ones~\cite{Sabrina2013Enforcing} based on selective encryption, or check the correctness of the ABS signature for the schemes~\cite{Ruj2012Privacy,Zhao2011Realizing} based on attribute-based cryptosystems.
In DVCS, the access control enforced at the single central server can easily be bypassed, as each peer may exchange the modification directly, moreover the central server may even not exist.

Here, we propose a distributed access control scheme (named Disac) for DVCS,  having the following properties:

(1) It achieves the access control for fully distributed VCS, as the set of shared data and the corresponding access control  policy is specified by the data owner and enforced at each entity.
Therefore, Disac avoids the unauthorized operations on the shared data stored at various peers, without breaking the  efficient and flexible cooperation of DVCS.

(2) It supports both the read and write access control.  For read, Attribute-based encryption (ABE) is adopted to ensure only the ones whose attributes satisfy the read policy can  perform the decryption successfully. For  write, the correct peers only accept the revision with the valid Attribute-based signature (ABS), which can only be generated by the entities whose attributes satisfy the write  policy.
Although the malicious entity may push the incorrect revisions to tamper the data, others can still extract the correct one from the  historical versions.

(3) It is easy to be integrated with existing DVCS.  Disac only needs to add two functions in  DVCS client, one is to  perform the ABE encryption and ABS signing before publishing the revision, the other is to invoke the ABS verification and ABE decryption after pulling the data.
\section{Background and related work}
\label{sec:backandrelatedwork}

In this section, we provide a brief description of DVCS, ABE and ABS, and introduce the related work. 

\subsection{Distributed version control system}

DVCS makes the cooperation on a project more efficient as each participant has a complete repository and modifies the data locally.
The cooperation is flexible as each entity is free to revise the data privately, choose the set of modifications to publish and modify 
based on any obtained shared version.
No single central server exists and the entire project is stored in multiple participants, thereby avoiding the single failure point.

Git is a typical DVCS that has been widely deployed. 
In Git, the repository stored in each entity does not store the version differences (i.e., patches) for version replacement, but records a snapshot of each version.
A new version is generated when the user adopts \verb+git commit+ to commit its batched modifications.
Each entity uses \verb+git clone+ or \verb+git fetch+ to obtain the repository (including the historical versions) from other peers,  employs \verb+git push+ to send its local repository to the remote one, executes \verb+git branch+ to create a new branch (i.e., an independent context) or to delete existing branches, adopts \verb+git checkout+ to switch to the specified branch, and uses \verb+git merge+ to merge two different versions.
These operations are  executed through Git client, which consists of working directory, index and head.
The working directory holds the actual data, index is the staging area, and head points to the last commit~\cite{git.com}.  

\subsection{ABE and ABS}

ABE is firstly proposed in~\cite{sahai2005fuzzy}, and divided into Key-Policy ABE (KP-ABE) and Ciphertext-Policy ABE (CP-ABE) in~\cite{goyal2006attribute}.  
In KP-ABE~\cite{goyal2006attribute}, the user's secret keys are associated with the access control policy.
While in CP-ABE~\cite{bethencourt2007ciphertext}, the user's secret keys are associated with their attributes and the access policy is contained in the ciphertext.
For the access control on cloud storage, CP-ABE provides better scalability, as the user only needs to maintain one secret key to decrypt the data with different access control policy.

CP-ABE~\footnote{ABE refers to the  CP-ABE  for simplicity hereafter.} consists of  four functions.
$ABE.Setup$ is invoked by the Attribute Authority (denoted as $AA$)  to generate the system public parameter $PP_r$ and the master key $MSK_r$.
For each user, $AA$ invokes $ABE.KeyGen$ with $MSK_r$ and the user's attributes, to generate the user's secret key $SK_r$.
For each data $d$ with the access control policy $AC_r$, any entity can invoke $ABE.Encrypt$ with the parameters $PP_r$, $AC_r$ and $d$ to generate the ciphertext (denoted as $C_d$).
However, only the entity $i$ whose attributes satisfy $AC_r$, can obtain the plaintext $d$ by invoking the $ABE.Decrypt$ with the public parameter $PP_r$, its secret key $SK_r^i$ and the ciphertext $C_d$. 

ABS is firstly proposed in~\cite{maji2011attribute}, to allow only the ones whose attributes satisfying the predefined access control policy can generate the valid signatures.
Similar to ABE, the $AA$ invokes $ABS.Setup$ to generate the  system public parameter $PP_w$ and the master key $MSK_w$, and generates the secret key $SK_w$ for each user by invoking $ABS.KeyGen$ with  $MSK_w$ and the user's attributes.
For each data $d$, there is a predefined access control policy $AC_w$, only the one whose attributes satisfy $AC_w$, can invoke $ABS.Sign$ with $PP_w$, $AC_w$, $SK_w$ and $d$ to generate the correct ABS signature.
Any entity checks the correctness of the ABS signature by invoking $ABS.Verify$ with the parameters $PP_w$, $AC_w$ and $d$.

\subsection{Related work}
The access control schemes on cloud storage can be applied in the CVCS, as only the central server stores the entire project and may perform the access control. 
These schemes introduce significant overhead to the entity who enforces  access control, which may be afforded by the central server but not each peer in DVCS.
For example, in the scheme~\cite{Ateniese2006Improved} based on proxy re-encryption, the entity needs to perform an independent re-encryption of each version for each delegated user, and in the schemes~\cite{Vimercati2007Over,Sabrina2013Enforcing} based on selective encryption, the entity needs to maintain the  key derivation graph dynamically which is not suitable for a larger number of cooperators. 
In Disac, each entity only needs to perform one ABE encryption and  ABS signature for one shared version to push.

Similar to~\cite{Ruj2012Privacy,Zhao2011Realizing}, Disac adopts ABE and ABS to protect the confidentiality  and integrity of the shared data.
However, these schemes~\cite{Ruj2012Privacy,Zhao2011Realizing} rely on a central server to check the revisions and only store the correct one, which is not suitable for DVCS.
LightCore~\cite{jiang2015lightcore} adopts the stream ciphers to provide  the confidentiality for real-time collaborative editing systems, and relies on the honest servers to ensure the data consistency.
Disac doesn't rely on a central server to maintain the correct revisions, and the revisions are checked by the peers.

Various schemes are proposed for securing Git. 
Gitolite~\cite{gitolite} provides the access control for Git when there exists a central server, and cannot be applied in fully distributed model.
Git-crypt~\cite{gitcrypt} encrypts the sensitive information in the public repository based git filter~\cite{gitfilter}. 
Git-remote-gcrypt~\cite{git-remote-gcrypt} encrypts the entire repository.
In Git-crypt and Git-remote-gcrypt, the key for decryption is distributed through GPG, which increases the overhead of the data owner for key management.
Similar to Git-crypt, Disac can also be integrated in Git through git filter, and provides the protection for both the partial or entire repository.
Moreover, the key management in Disac is much easier as the data owner only needs to perform one ABE encryption and ABS signing for various users.
\section{System Overview}
\begin{figure}[b]
\centering
\setlength{\abovecaptionskip}{-0cm}
\centerline{\epsfig{figure=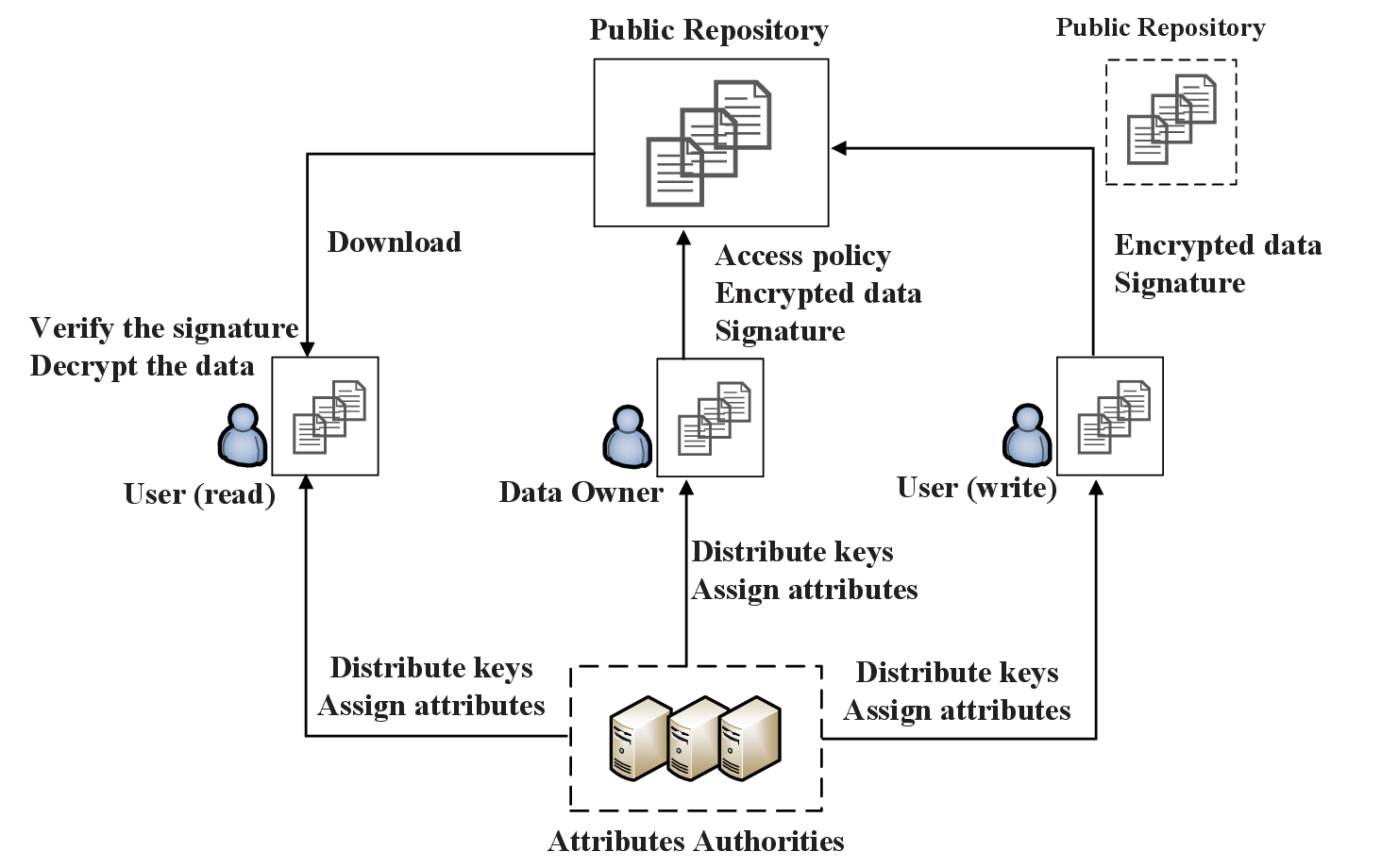,width=8.5cm}}
\caption{The architecture of Disac.}\label{fig:systemmodel}
\end{figure}

As shown in Figure~\ref{fig:systemmodel}, Disac consists of multiple public repositories, Attribute Authorities ($AA$), and the users with different privileges.
The data owner is the one who creates the original version of the data, and has the full control over the data, no matter which repositories the data is in. 
That is, the data owner is the only  one to specify and modify the  users' privileges on the  data.
The public repositories provide the stored data to all users, and store the revisions pushed by any entity.

Each $AA$ manages attributes of its corresponding field, establishes the system public parameter ($PP_r$ and $PP_w$) and the master key ($MSK_r$ and $MSK_w$) for the ABE and ABS at the setup phase.
For each user $i$, $AA$ assigns the attributes, generates and distributes the corresponding secret keys ($SK_r^i$ and $SK_w^i$) for the read and write operation respectively.  

The data owner specifies the access control policies for read and write permission (denoted as $AC_r$ and $AC_w$) by defining the corresponding attribute structures, adds $AC_r$ and $AC_w$ to the metadata of the data, encrypts the data using ABE and generates the signature of the data with ABS before uploading it to the public repository. The metadata is not encrypted, and can only be modified by the data owner.

All users may download  data from the public repository.
However, only  users whose attributes satisfy $AC_r$, can obtain the plaintext through ABE decryption. 
To update the data, the users whose attributes meet $AC_w$ have to generate an ABS signature of the revision.
Only when the signature is valid according to $PP_w$ and $AC_w$, the new version is accepted.

\noindent{\textbf{Threat Model.}} In Disac, we assume each $AA$ is trusted and well protected, never leaks $MSK_r$, $MSK_w$, and the secret keys of each user. 
The entities hosting the public repositories  are honest-but-curious, they store each version of the data and provide the requested data correctly, however, they may want to perform  unauthorized read or write operations.  
The correct user $i$ accepts the received version only when the corresponding ABS signature is valid,  decrypts the data with $SK_r^i$, generates the signature of the modified data with $SK_w^i$, and never leaks $SK_r^i$ and $SK_w^i$ or performs  ABE encryption with a policy different from the one specified by the data owner.
However,  malicious users may behave arbitrarily, for example, attempting to infer the plaintext of unauthorized data, upload  invalid modifications   and perform DDoS attacks.
Moreover, malicious users may collude for more privileges.
\section{Distributed Access Control Scheme}
\label{sec:Disac}

In Disac, the data stored in the DVCS is in the form of [$meta$, $Cipher_d$, $Sig_d$], where $meta$ includes $AC_r$, $AC_w$ and the unique identifier of the data and data owner, $Cipher_d$ contains the ciphertext of the data $d$, while $Sig_d$ is the ABS signature of $Cipher_d$.
The public repositories manage the versions of data [$meta$, $Cipher_d$, $Sig_d$].

\begin{figure}[t]
 \centering
 \centerline{\epsfig{figure=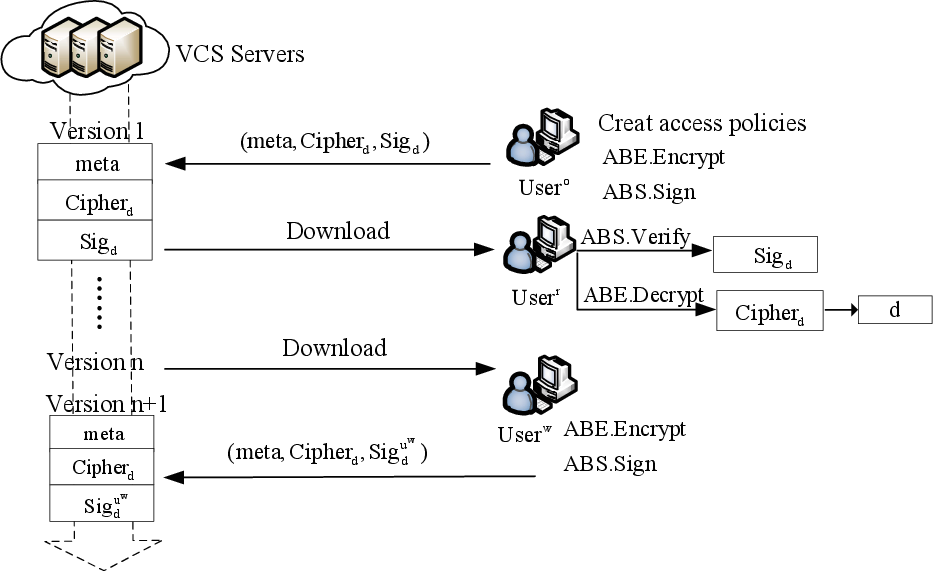,width=8.5cm}}
\label{2}
\caption{The data processing in Disac.}\label{fig:disacprocess}
\vspace{-5 mm}
\end{figure}

Disac supports  data creation, reading, modification and deletion, and can be  integrated with the clients of DVCS  to provide all the common DVCS functions. To make the description clear, as demonstrated in Figure~\ref{fig:disacprocess}, we first provide the process details of Disac, under the cases that all the users obtain the data from one public repository, where user $u^o$ creates the data $d$, user $u^r$ wants to get the latest valid version of $d$, while the user $u^w$ uploads the modification on $d$ to the public repository. The process that a user operates on the data from multiple repositories is provided in Section~\ref{subsec:multiRepo}.

\subsection{Data processing in Disac}
\label{subsec:dataprocess}

\noindent{\textbf{Data creation.}} The data owner $u^o$ processes as follows:

(1) Firstly, $u^o$ sets access policies $AC_{r}^{d}$ and $AC_{w}^d$ for the read and write permission respectively, and constructs the metadata using $AC_{r}^{d}$, $AC_{w}^d$, with the identifier of $d$ and $u^o$. 

(2) Secondly, $u^o$ generates a random symmetric key $DK$, encrypts $d$ with $DK$ to obtain the ciphertext of $d$ (denoted as $C_d$), and adopts the ABE to encrypt $DK$ for the ciphertext of $DK$ (denoted as $C_{DK}$), that is $C_{DK}=ABE.Encrypt(PP_r,AC_r^d,DK)$. Therefore, the $Cipher_d$ is in the form of [$C_d$, $C_{DK}$].

(3) Then, $u^o$ adopts ABS to generate $Sig_d$, that is, $Sig_d=ABS.Sign(PP_w,AC_w^d,SK_w^o,Hash(Cipher_d))$, where $SK_w^o$ is the write secret key of $u^o$, and $Hash(Cipher_d)$ is the digest of $Cipher_d$.

(4) Finally, $u^o$ commits [$Cipher_d$,$Sig_d$] with the metadata to its local repository, and pushes it to the remote ones.

\noindent{\textbf{Data reading.}} The user $u^r$ may obtain any version of $d$ from the public repository. Here, we use the reading of the latest valid version as an example. After obtaining the latest version of [$Cipher_d,Sig_d$], $u^r$ checks the signature $Sig_d$ by invoking $ABS.Verify(PP_w,AC_w^d, Hash(Cipher_d), Sig_d)$. The version is valid only when the ABS signature is correct.
Then, $u^r$ extracts $C_{DK}$ and $C_d$ from $Cipher_d$, invokes $ABE.Decrypt(PP_r,SK_r^{u^r}, C_{DK})$ to obtain $DK$, and uses $DK$ to decrypt $C_d$ to obtain the plaintext of $d$.
If $u^r$ doesn't have the read permission of $d$, it fails to obtain correct $DK$ from the ABE decryption.
If the $Sig_d$ is incorrect, $u^r$ gets the previous version of $d$ from the repository, and performs the ABS verification again until it obtains a latest valid version.

\noindent{\textbf{Data modification.}}
To make the description clear, we firstly provide the process for  $u^w$ to generate the new version of $d$ based on the latest version. In this case, the process is the same as the step 2 to 4 for the data owner to upload the initial version of the data.

However, $u^w$ may perform the modification based on a version $ver_m$ different from the latest version stored in the public repository, which means a conflict exists. 
In this case, $u^w$ downloads all versions that are later than $ver_m$, and checks whether these versions are valid.
If the latest valid version $ver_n$ is larger than $ver_m$, $u^w$ obtains the plaintext of the version $ver_n$ of $d$,~\footnote{To make the revision useful, the modifiers also have the read permission.} transfers its operations on the version $ver_m$ to the ones on $ver_n$, and uploads the modification to the public repository as described above.
Otherwise, $u^w$ directly uploads its local modification to generate a new version $ver_n+1$.

\noindent{\textbf{Data Deletion.}} 
In Disac, the users who have the write permission, may delete $d$.
However, in this case, the same as existing DVCS, the authorized users can still download all the previous versions of $d$, while the access control specified in the metadata still prevents the unauthorized read and write.  

\subsection{Access control management in Disac}
\label{subsec:dis}
The process of data creation, reading, modification and deletion may be integrated in DVCS clients, to make it transparent to  users for complex operations.

\noindent{\textbf{Access control policy management.}}
In Disac, the data owner modifies the access control policy of the data $d$ as follows:

(1) To revoke the read permission of the users (${\mathbb{U}^{r-}}$), the data owner needs to modify  $AC_r^d$ in the metadata.
Then the following data modifiers perform the ABE encryption of the newly generated $DK$ with the new $AC_r^d$, which ensures that the users in ${\mathbb{U}^{r-}}$ fail to obtain the further versions of the data.

(2) To grant read permission to the users (${\mathbb{U}^{r+}}$), the data owner modifies the $AC_r^d$ in the metadata by adding attributes, which allows these users to read the further versions of the data. For the historical versions, the data owner firstly determines the version numbers (denoted as $\mathbb{V}^h$) to be provided to ${\mathbb{U}^{r+}}$;
obtains $DK^i$ (the $DK$ for version $v^i$) by decrypting $C_{DK}^i$ (the $C_{DK}$ for version $v^i$) for each version number $v^i$ in $\mathbb{V}^h$;
then, uses the new $AC_r^d$ to invoke ABE to generate the encryption of $\{(v^i, DK^i) | v^i \in \mathbb{V}^h\}$;
and sets the ciphertext as a special domain in the data uploaded to the public repository, which allows the users in ${\mathbb{U}^{r+}}$ to decrypt the versions (in $\mathbb{V}^h$) of the data obtained from the public repository.

(3) To revoke the write permission of a set of users (${\mathbb{U}^{w-}}$), 
or grant the write permission to users in ${\mathbb{U}^{w+}}$, the data owner modifies $AC_w^d$ and uploads it to the public repository, which generates a new version (denoted as $v^w$).
Then, for the users in  ${\mathbb{U}^{w-}}$, their modifications on $d$ before $v^w$ are still accepted while the ones after $v^w$ will be rejected by others;
for the users in  ${\mathbb{U}^{w+}}$, their modification on $d$ after $v^w$ are accepted while the ones before $v^w$ are invalid.

\noindent{\textbf{Attribute management.}}
In Disac, the access policy is based on the attribute which is managed by the trusted $AA$.  In the practical DVCS, the data may be shared and managed  with the users whose attributes are managed with a different authority. Disac may apply the multi-authority ABE and ABS instead of the traditional ABE and ABS.
Moreover, some attributes may need to be revoked for a user, the authority may adopt the fine-grained attribute revocation scheme or update the secret keys directly for revocation.

\noindent{\textbf{Metadata Protection.}}
In Disac, the metadata contains $AC_{r}^{d}$, $AC_{w}^d$, the identifiers of the data ($ID^d$) and data owner ($ID^o$).
As described above, the metadata can only be modified by $ID^o$, to avoid the permissions  granted  to others  without the control of $ID^o$.
To achieve this, $ID^o$ needs to generate the signature ($Sig^{meta}$) of [$AC_{r}^{d}$, $AC_{w}^d$, $ID^d$, $ID^o$] using its private key while the corresponding public key is combined with $ID^o$ through a valid  X.509 certificate. 
To create the data $ID^d$, $ID^o$ uploads [$AC_{r}^{d}$, $AC_{w}^d$, $ID^d$, $ID^o$, $Sig^{meta}$], which is the metadata for the first version of $ID^d$.
To modify the metadata, $ID^o$ generates $Sig^{meta'}$ of $[AC_{r}^{d'}, AC_{w}^{d'}, ID^d, ID^{o'}]$.
Others accept $AC_{r}^{d'}$ and $AC_{w}^{d'}$ as the new access policies for $ID^d$ only if $Sig^{meta'}$ is valid and $ID^{o'}$ is the same as $ID^o$ in the metadata of the first version of $ID^d$.
  
\noindent{\textbf{Access control granularity.}}
It is common that  only several files need to be  protected in the project.
Therefore, in Disac, the data owner is the creator of each file instead of the whole project, which achieves the fine-grained access control.

\subsection{Access control in fully distributed VCS}
\label{subsec:multiRepo}
In this section, we demonstrate the process when the users operate the data from multiple public repositories. 
The same as in Section~\ref{subsec:dataprocess}, each user adopts ABS to ensure the integrity of the data and ABE for the confidentiality.

The main difference in fully distributed VCS, is that no blessed repository exists to maintain the reference version, and  each peer provides a public repository which introduces conflicts. The conflict exists if the obtained revision sequences are inconsistent in different repositories. 
If the different revision sequences may be merged, the user performs the merge directly.
Otherwise, the user creates one independent branch for each revision sequence, manually chooses the one to work on, and switches to the other when necessary. 
Similar to blockchain, the entities will reach an agreement on the  revision sequence  while the different revision sequences still exist in the historical versions.

\subsection{Disac applied in DVCS}
\label{subsec:example}
Disac can be easily applied to DVCS. Here, we use Git as an example. In Disac, the Git client only needs to add the process of access control before  uploading the data in the head to the remote repository and fetching the data from the remote repository.
That is, Disac only needs to modify the process of \verb+git clone+, \verb+git fetch+ and \verb+git commit+. 
The other operations are the same as the original Git.

The user adopts  \verb+git clone+ or \verb+git fetch+ to download all history for the specified repository.
In Disac, each downloaded version of data $ID^d$ is  a pair of $(ID^d, [Cipher_d,Sig_d])$ and [$AC_{r}^d$, $AC_{w}^d$, $ID^d$, $ID^o$, $Sig^{meta}$].  
Before accepting this version, the Git client processes as follows:
firstly, it checks  the correctness of $Sig^{meta}$ and the consistency of $ID^o$ with the one in the first version, and
only when both checks are passed, the metadata  is correct;
secondly, it adopts ABS verification with the parameters $AC_{w}^d$ and $PP_{w}$ to verify $Sig_d$;
then, it uses ABE decryption with $AC_{r}^d$ and $PP_{r}$ to decrypt $C_{DK}$ in $Cipher_d$ for  $DK$;
finally, it uses $DK$ to decrypt $C_d$ in $Cipher_d$ for the plaintext of $ID^d$.
If the metadata or $Sig_d$ is incorrect, this version is rejected by the Git client which prompts the corresponding warning information.
For a version whose metadata and signature are both valid, the Git client displays the plaintext in the user's working directory and allows it to be modified or deleted, thereby preserving the normal Git workflow.

The \verb+git commit+ is used to generate a new version locally. 
Firstly, the Git client uses the random $DK$ to encrypt the modification, adopts ABE to encrypt $DK$ with the parameters  $AC_{r}^d$ and $PP_{r}$, generates the ABS signature of $Cipher_d$ with the parameter $PP_{w}$ and its secret key $SK_w^{u^w}$, and saves  $(ID^d, [Cipher_d,Sig_d])$ with unmodified metadata to the repository. Then $u^w$ can use \verb+git push+ to upload the local commits to the remote repository.
However, conflicts may exist due to the inconsistency between the user's local versions and the remote one.
As the data in the user's working directory is in plaintext, the conflict process  in Disac is the same as the one in the original Git, that is, $u^w$ uses \verb+git fetch+ to obtain the correct versions in the remote repository, tries to auto-merge changes with \verb+git merge+, and performs the manual merges when necessary.

When $u^w$ creates a new file ($ID^f$), its Git client will construct the metadata of $ID^f$, submit to local repository and upload it to the remote repository.  
The process to change the access control policy on $ID^f$ is the same as the creation of $ID^f$, the Git client of $u^w$ uploads the newly generated metadata.
If the \verb+git rm+ operation is included in the local commits, the Git client  uploads  $(ID^d, [NULL,Sig_d])$ to complete the data deletion, where $Sig_d$ is the ABS signature of \verb+git rm+ operation (including the parameters).

The computation and storage overhead introduced by Disac is modest. 
The data (including access policy) creation and modification require one ABE encryption of $DK$, symmetric encryption and ABS signature. 
Data reading needs one ABS verification, ABE decryption for $DK$, symmetric decryption and signature verification, while the data deletion only needs one ABS signature. 
The metadata modification (including creation) needs a signature generation ($Sig^{meta}$). 
For each version, Disac requires additional storage for $AC_r^d$, $AC_w^d$, $C_{DK}$, $Sig_d$ and $Sig^{meta}$. 
Moreover, granting read permission of historical versions to ${\mathbb{U}^{r+}}$, requires extra computation and storage for the ABE encryption of the corresponding $DKs$.
\vspace{-3 mm}
\section{SECURITY ANALYSIS}
\label{sec:analysis}
In this section, we provide a brief security analysis of Disac. In Disac, any valid modification on the access control policy should contain the signature of the data owner, which will be verified by the other users. The modification is accepted only when the users determine that the generator of the signature corresponds to the one specified in the metadata of the initial version. Therefore, as only the data owner has the corresponding private key to generate the correct signature, the metadata can only be modified by the data owner.

In Disac,  correct users never leak their  secret keys $SK_r$ and $SK_w$ or perform the ABE or ABS with the incorrect policy, which ensures the illegal operations of unauthorized entities will not succeed.
For data $d$, the data owner and authorized modifiers upload [$C_d$, $C_{DK}$] where $C_d$ is the encryption of $d$ using the key $DK$, and $DK$ is encrypted by ABE with $AC_{r}^{d}$ specified in the metadata.
Any unauthorized entity, whose attributes do not satisfy $AC_{r}^{d}$, cannot obtain $DK$ from $C_{DK}$ successfully.
The correct entities accept the modification of $d$ only when the corresponding ABS signature $Sig_d$ is valid,
while the unauthorized entity whose attributes do not meet the $AC_{w}^{d}$ fails to generate the correct $Sig_d$.

The malicious users may attempt to collude for more read or write 
permissions. 
However, it's the basic requirement of ABE and ABS, that the users cannot collude to pool their attributes~\cite{goyal2006attribute,maji2011attribute}, which avoids the colluded users completing operations outside the permissions.

The user whose permissions are revoked becomes unauthorized, since the data owner uploads the version $v^r$  in which the metadata specifies new access control policy.
For read permission, the revoked users can obtain the previous versions ($< v^r$), but fail to obtain the further revisions ($\ge v^r$), as their secret key $SK_r$ cannot be used to perform ABE decryption for $DK$ which is different for each version.
For write permission, the  previous versions ($< v^r$) of the revoked users are still accepted by others, but the further revisions ($\ge v^r$) are not accepted, as the ABS signatures generated using their secret keys $SK_w$ are not valid with the new $AC_w^d$.

The malicious entity may perform DDoS attacks, by uploading the invalid modification.
Existing DDoS mitigation (e.g., limiting the uploading frequency for each IP address) is still useful in Disac. 
As Disac supports the fully distributed VCS, the difficulty of DDoS is larger than that in the central VCS, as no single server exists.
The overhead of the correct clients is modest as they only process the  versions with correct ABS signatures, and drop the invalid one directly.
\section{CONCLUSIONS}

In this paper, we propose Disac, a distributed access control scheme for fully distributed VCS, to avoid  unauthorized read or write on the sensitive data in the public repositories.
In Disac, the data owner has full control over the shared data, and specifies the access control policy based on the attributes.
The policy is enforced at each user, as only the authorized users can invoke the ABE decryption successfully to obtain the plaintext, while only the revisions made by the authorized users have the correct  ABS  signatures to be accepted by correct users.
Moreover, the integration of Disac with Git demonstrates that Disac is easy to deploy.

\bibliographystyle{IEEEbib}
\bibliography{ref}

\end{document}